# Domain imaging and domain wall propagation in (Ga,Mn)As thin films with tensile strain

K. Y. Wang

*Hitachi Cambridge Laboratory, Cambridge CB3 0HE, United Kingdom*

A. W. Rushforth, V. A. Grant, R. P. Campion, K. W. Edmonds, C. R. Staddon, C. T. Foxon, and B. L. Gallagher

*School of Physics and Astronomy, University of Nottingham, Nottingham NG7 2RD, United Kingdom*

J. Wunderlich, D. A. Williams

*Hitachi Cambridge Laboratory, Cambridge CB3 0HE, United Kingdom 2RD, United Kingdom*

*Abstract*

We have performed spatially resolved Polar Magneto-Optical Kerr Effect Microscopy measurements on as-grown and annealed $Ga_{0.95}Mn_{0.05}As$ thin films with tensile strain. We find that the films exhibit very strong perpendicular magnetic anisotropy which is increased upon annealing. During magnetic reversal, the domain walls propagate along the direction of surface ripples for the as-grown sample at low temperatures and along the [110] direction for the annealed sample. This indicates that the magnetic domain pattern during reversal is determined by a combination of magnetocrystalline anisotropy and a distribution of pinning sites along the surface ripples that can be altered by annealing. These mechanisms could lead to an effective method of controlling domain wall propagation.

The (III,Mn)V ferromagnetic semiconductor family has attracted much attention for its potential applications in spintronics, in which, for examples, logic and memory operations may be integrated on a single device [1,2]. (Ga,Mn)As, with Curie temperature as high as 173 K [3], is one of the most promising candidates for such applications. Understanding and controlling the magnetic properties, including domain structure and domain wall propagation in these materials is important for realizing and manipulating spintronics devices.

(Ga,Mn)As epilayers grown on a relaxed (001) (In,Ga)As buffer layer experience a tensile strain due to the difference in the lattice constant in each layer. Under these conditions the magnetic easy axis is known to be perpendicular to the plane [4] in agreement with theoretical predictions [1]. Magnetic domain patterns in (Ga,Mn)As with perpendicular magnetic anisotropy have been studied previously using Scanning Hall Probe Microscopy [5] and Polar Magneto-Optical Kerr Effect Microscopy (PMOKM) [6-8].Ref. 5 showed that domains formed a stripe pattern with a typical width of a few microns at low temperatures. The stripe domains were aligned nearly parallel to the <110> directions. This was attributed to the magnetocrystalline anisotropy of the films. In this paper we employ PMOKM to study magnetic domains over an area (150 μm ×150 μm ) much larger than reported in Ref. [5] (7.3 μm × 7.3 μm), and show a clear correlation between magnetic and topographic features. These images reveal that, at the lowest temperatures the magnetic stripe domains are aligned along the topological crosshatch pattern, known to form along the <110> directions due to mismatch dislocations in the buffer layer [ 9, 10]. This indicates that an additional mechanism, besides the magnetocrystalline anisotropy, is responsible for determining the size and direction of the magnetic stripe domains during magnetization reversal at low temperatures.

The epilayer structure was grown by Molecular Beam Epitaxy using a modified Varian GEN-II system [11]. The growth sequence consisted of a semi-insulating GaAs(001) substrate onto which was deposited a high temperature GaAs buffer layer at 580°C. This was followed by a 580nm $In_{0.15}Ga_{0.85}As$ layer grown at 500°C and a 25nm $Ga_{0.95}Mn_{0.05}As$ layer grown at 255°C. Post-growth annealing was performed in air at 190 $^0$C for 120 hours, which is an established procedure for increasing the $T_C$ of (Ga,Mn)As thin films [12]. X-ray diffraction measurements were performed using a Philips X'Pert Materials Research Diffractometer. By analysing both symmetric and asymmetric reciprocal space maps taken in the two <110> directions [13], the degree of relaxation of the (In,Ga)As layer was estimated to be 70% in both directions, consistent with previous work [14] which showed that above a certain thickness, the (In,Ga)As layer can support itself without further relaxation. The wafer was cleaved into separate pieces for PMOKM measurements and SQUID magnetometry. The PMOKM images were obtained using a commercial system with a high pressure Hg lamp and a high resolution CCD camera giving a spatial resolution of 1 μm. The SQUID measurements were carried out using a commercial Quantum Design system.

Figures 1(a) and (c) show magnetic hysteresis loops measured using PMOKM for the as-grown and annealed film respectively. The Kerr rotation angle, averaged over the image area, is proportional to the component of the magnetization pointing perpendicular to the plane of the film. Both samples show nearly square hysteresis loops, indicating that the magnetic easy axis is perpendicular to the plane over the whole temperature range below the Curie temperature ($T_C$). This is also confirmed by the temperature dependence of the remnant magnetization (Figs1 (b) and (d)), which is fit well by a Brillouin function with S=5/2. Low temperature annealing increases $T_C$ from 66K to 137K and increases the magnetization, as observed by the increase in the Kerr rotation angle and confirmed by SQUID magnetometry. This is expected since the annealing process removes interstitial Mn ions [15] which act as double donors and couple antiferromagnetically to the substitutional Mn ions. The removal of these impurities also leads to increase of the tensile strain and increase of hole density [16], thus resulting in a stronger perpendicular magnetic anisotropy [17]. Evidence for this is the fact that, for the annealed sample, the hysteresis loops are more square. The coercive field is also reduced, possibly due to a reduction in the number of pinning sites or the pinning energy, as will be discussed later. We note that, for the as–grown sample, the $T_C$ measured by SQUID is slightly smaller than that measured by PMOKM, probably due to small variations between the two pieces of the wafer used for each measurement. There is also a small in-plane component of the remnant magnetization measured in the as-grown sample. This is not observed in the annealed sample.

Figures 2(a) and (b) show Atomic Force Microscopy (AFM) images of the as-grown sample. These images reveal ripples in the surface running along the <110> directions to form a crosshatch pattern. The ridges running along the [110] directions have an average separation of 1 μm and those running along the [1-10] direction have an average separation of 0.5 μm. The ridges are typically ~10nm in height. These ridges are known to occur during the growth of the (In,Ga)As buffer layer, due to the formation of strain-relieving misfit dislocations [8]. PMOKM images at 6K of an as-grown sample from the same wafer are shown in Figs. 2(c) and (d). Initially, the film is saturated with negative magnetic field of -300 Oe, which is much larger than the coercive field. The field is then swept to 145 Oe, just less than the coercive field. PMOKM images are then captured at a rate of 15 frames per second. The domain images captured at 0s and 1.3s are shown in Figures 2(c) and (d), respectively. These images reveal that the magnetization reversal proceeds through the nucleation and propagation of magnetic domains (shown by dark features in Figs 2(c) and (d)). The domains propagate along the ripples, predominantly in the [110] direction, increasing in length, but not in width. After 12s a stable state is obtained indicating that thermal excitation and the small applied field do not provide enough energy to overcome the local barriers to domain wall motion. The magnetic domains form stripes of typical width 2-3 μm, a similar length scale to the separation of the ridges. At higher temperatures, the magnetic reversal has two main differences to that at low temperatures. Firstly, full reversal occurs in a time period of tens of seconds. Secondly, the domain walls propagate along random directions forming a cauliflower-like domain pattern during the magnetic reversal and the average domain size increases with increasing temperature. This is due to a lowering of the in-plane magnetic anisotropy barriers and a more effective thermal activation of domain wall motion. This behaviour at high temperatures is similar to that observed in a previous PMOKM study of tensile strained (Ga,Mn)As films [6].

Figure 3 shows successive snapshots of magnetic domain reversal at T =90 K in the (Ga,Mn)As film after annealing. Similar images are observed over the whole temperature range for the annealed sample. They differ from the images obtained for the as-grown sample in that the reversed domains are typically much larger (several tens of microns) during the reversal. The domain walls align along the [1-10] direction and propagate rapidly along the [110] axis between pinning sites, until the magnetization is almost fully reversed with only a few unreversed stripe domains remaining. The width of the unreversed stripe domains is a few microns, which is much wider than the typical domain wall width (~15 nm) in (Ga,Mn)As [18]. At lower temperatures similar behaviour is observed.

A previous study [5] of the magnetic domain structure of (Ga,Mn)As with perpendicular magnetic anisotropy found that the magnetic domains formed a stripe pattern and the width of the stripe domains did not depend strongly on temperature until temperatures close to $T_C$. Also the orientation of the stripes rotated from the <110> directions to the <100> directions as temperature increased. For our as-grown sample stripe domains are only observed at T<20K and are always oriented along the <110> directions. This suggests that the formation of stripe domains in our sample arises because the domain walls are pinned at features which are themselves aligned along the <110> directions. Above about 20K, thermal excitation is sufficient to overcome the energy barriers at these pinning sites and the stripe domain pattern is not observed. Recently, it has been proposed [10] that it may be energetically favourable for the interstitial Mn atoms to be concentrated at the ridges of the crosshatch pattern. Alternatively, the misfit dislocations may provide more favourable sites for Mn interstitials in the trough regions. In either case, a distribution of the density of Mn interstitials along the direction of the surface ripples may provide a network of pinning sites that cause the domain walls to be preferentially aligned along the <110> directions. Annealing removes the interstitials, thereby reducing the density of such pinning sites and the stripe domain pattern is no longer observed at any temperature. However, the fact that the domain walls are always aligned with the [110] direction after annealing may be due to residual pinning sites or may be due to the weak magnetocrystalline anisotropy [5].

We have studied the magnetic properties of a tensile strained $Ga_{0.95}Mn_{0.05}As$ thin film by using SQUID and PMOKM. Both the as-grown and annealed samples show very strong uniaxial anisotropy with easy axis along [001] direction and the Curie temperature up to $137 \pm 2K$ for the annealed sample. During the magnetic reversal, for the as-grown sample the domain walls tend to propagate along the direction of the surface ripples at low temperatures, which indicates that an additional mechanism to the magnetocrystalline anisotropy is responsible for the magnetic domain pattern and the propagation of magnetic domain walls. This may be the presence of pinning sites caused by Mn interstitials distributed along the crosshatch pattern. This mechanism and the strongly anisotropic domain wall motion found after annealing have implications for studies of current-driven domain wall motion in ferromagnetic semiconductor devices and may provide an effective method for control of the formation of magnetic domains and propagation of domain walls.

*ACKNOWLEDGEMENT*

This project was supported by EC sixth framework grant: FP6-IST-015728 and EPSRC (GR/S81407/01).

*Figure Captions:*

Fig.1 (a) and (c): Magneto-optical Kerr rotation hysteresis loops obtained from PMOKM for (a) as-grown and (c) annealed $Ga_{0.95}Mn_{0.05}As$ thin film at different temperatures. (b) and (d) Temperature dependent of normalized magnetization from PMOKM (solid symbols) and SQUID (open symbols) for (b) the as-grown and (d) annealed $Ga_{0.95}Mn_{0.05}As$ thin film. The line is the Brillouin function with S=5/2, using the Tc obtained from the PMOKM measurements.

Fig. 2 (a) Topology and (b) derivative AFM image for the as-grown $Ga_{0.95}Mn_{0.05}As$ thin film; successive snapshots of the magnetization reversal at 6K for the as-grown $Ga_{0.95}Mn_{0.05}As$ thin film at H = 145 Oe, (c) 0 s; (d) 1.3 s respectively.

Fig.3 Successive PMOKM snapshots of the magnetic domain pattern during the magnetization reversal at 90K for the annealed $Ga_{0.95}Mn_{0.05}As$ thin film at H = 23.7 Oe, (a) 5 s; (b) 5.6 s; (c)7.3 s; (d) 9.23 s; and (e) 9.37s, respectively. Arrow in (e) indicates a persistent residual stripe domain.

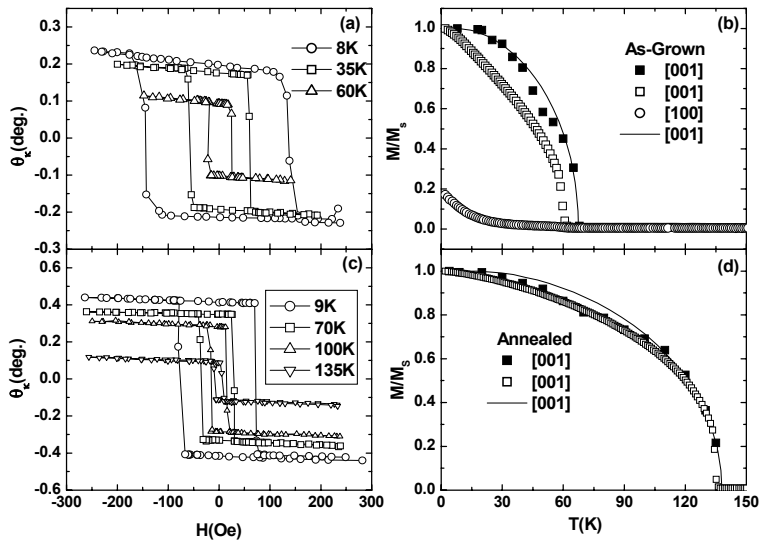

*Fig.1 K. Y. Wang et al.*

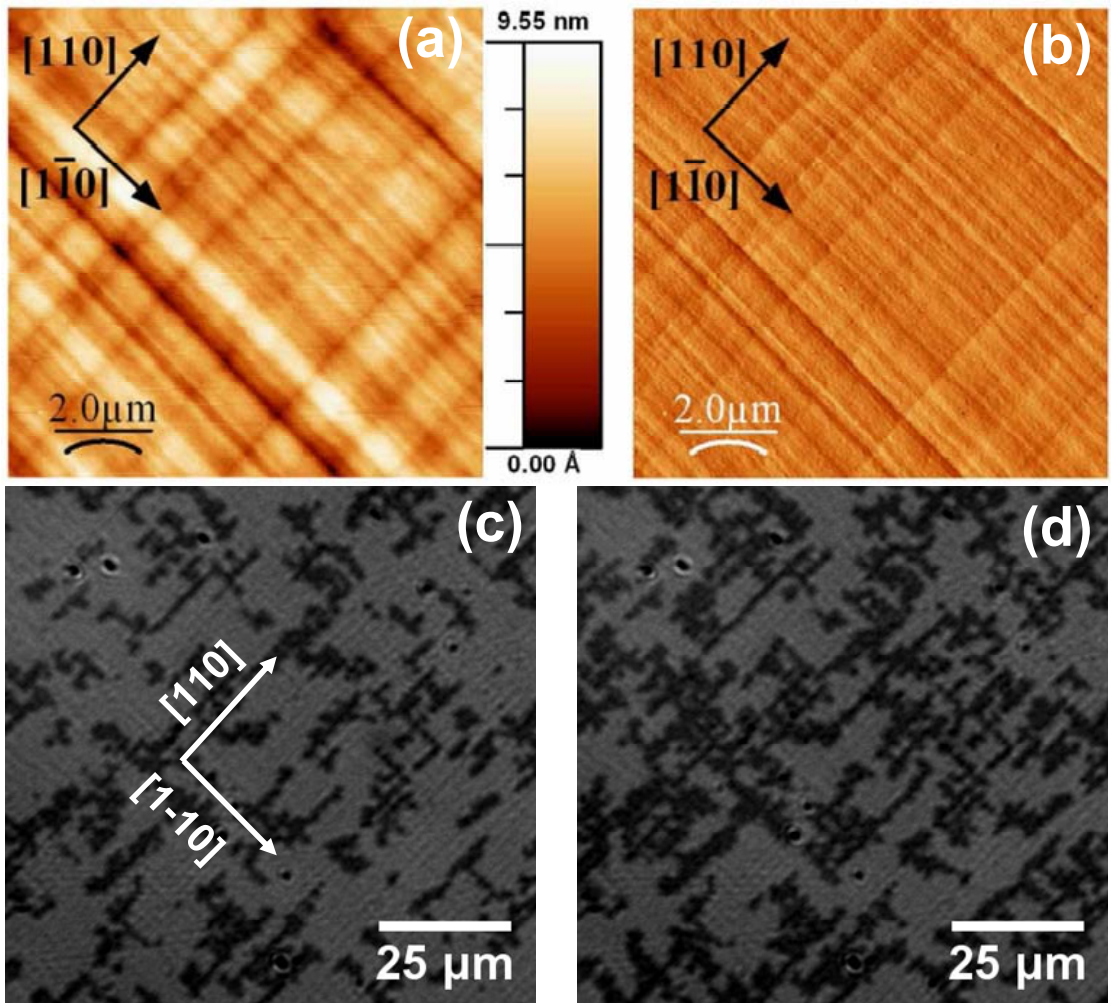

*Figure 2 K. Y. Wang et al.*

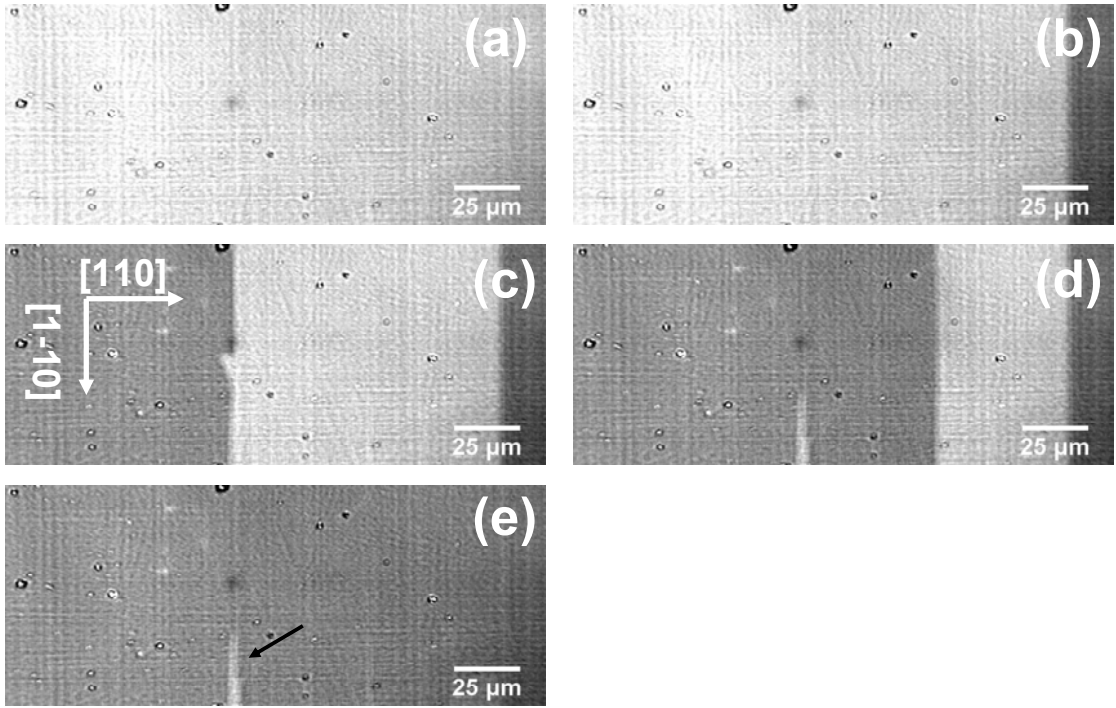

*Figure 3 K. Y. Wang et al.*